\documentclass[a4paper,twocolumn,prl,showpacs]{revtex4-1}
\usepackage{amsmath}
\usepackage{amssymb}

\usepackage{graphicx}
\usepackage{epsfig}
\usepackage{ulem}
\usepackage{color}

\begin{document}

\title{Suppression of Anderson localization of light and Brewster anomalies in disordered superlattices containing a dispersive metamaterial}

\author{D. Mogilevtsev$^{1,2}$, F. A. Pinheiro$^{3}$, R. R. dos Santos$^{3}$, S. B. Cavalcanti$^{4}$, and L. E. Oliveira$^{2}$}
\address{$^{1}$Institute of Physics, NASB, F. Skarina Ave. 68, Minsk, 220072, Belarus\\
$^{2}$Instituto de F\'{i}sica, UNICAMP, CP 6165, Campinas-SP,
13083-970, Brazil\\
$^{3}$Instituto de F\'{i}sica, Universidade Federal do Rio de Janeiro, Rio de Janeiro-RJ, 21941-972, Brazil\\
$^{4}$Instituto de F\'{\i}sica, Universidade Federal de Alagoas, Macei\'{o}-AL, 57072-970, Brazil}
\begin{abstract}
Light propagation through 1D disordered structures composed of alternating layers, with random thicknesses, of air and a dispersive metamaterial is theoretically investigated. Both normal and oblique incidences are considered. By means of numerical simulations and an analytical theory, we have established that Anderson localization of light may be suppressed: (i) in the long wavelength limit, for a finite angle of incidence which depends on the parameters of the dispersive metamaterial;
(ii) for isolated frequencies and for specific angles of incidence, corresponding to Brewster anomalies in both positive- and negative-refraction regimes of the dispersive metamaterial.
These results suggest that Anderson localization of light could be explored to control and tune light propagation in disordered metamaterials.
\end{abstract}

\date{\today}
\maketitle

In the last decade or so, advances in metamaterials have allowed the development of unusual optical properties, with no counterpart in natural media, opening up new frontiers in photonics. For instance, metamaterials exhibit negative refraction~\cite{Smith00}, resolve images beyond the diffraction limit~\cite{smith2004}, exhibit optical magnetism~\cite{enkrich2005,cai2007}, act as an electromagnetic cloak~\cite{pendry2006,leonhardt2006} and yield slow light propagation~\cite{OPN}.

Notwithstanding the wide range of new physical phenomena unveiled sofar, there are fundamental issues that are still not fully understood. Anderson localization (AL) of light in disordered metamaterials is certainly one of them. The concept of AL was originally conceived in the context of condensed matter physics as the vanishing of electronic diffusion in disordered systems~\cite{Anderson58}. Being an interference wave phenomenon, this concept has been extended to light, acoustic waves, and even Bose-Einstein--condensed matter waves~\cite{bartptoday}. In the case of light, its vector character, together with the recent development of metamaterials with unusual electromagnetic properties such as the possibility of optical magnetism, should lead to interesting particularities in AL. Indeed, the vector character of light is at the origin of a polarization-induced anomalous delocalization effect in 1D disordered systems~\cite{sipe1988}, for which the vast majority of states is localized~\cite{sheng}.
The presence of metamaterials in a 1D superlattice also leads to light delocalization in the long-wavelength limit~\cite{asatryan}. However, given that in Ref.~\cite{asatryan} the metamaterial was treated as non-dispersive (which leads to negative electromagnetic energy density~\cite{Ramakrishna,soukoulisbook}) and that only normal incidence was considered, a more thorough investigation is clearly in order.

The aim of the present Letter is to discuss light propagation through randomly perturbed 1D photonic heterostructures composed of alternating layers of non-dispersive right-handed (RH) materials (labeled A) and dispersive left-handed (LH) metamaterials (label M). We will establish that allowing for dispersion of both dielectric permittivity and magnetic permeability of the M layers, together with oblique incidence leads to novel features of AL in 1D. In particular, we find unexpected extended states at Brewster angles for both TM and TE configurations, a high transmission peak at the very edge of a bandgap, and light delocalization in the low-frequency regime at finite angles of incidence.

We model our system as a stack of alternating layers of air ($\epsilon_A=\mu_A=1$) and of a Drude-like metamaterial, with responses for the dielectric permittivity and magnetic permeability of the M layer given as~\cite{epsilon-mu}
\begin{equation}
\epsilon_M(\omega)=\epsilon_0- \frac{\omega_e^2}{\omega^2},\
\mu_M(\omega)=\mu_0- \frac{\omega_m^2}{\omega^2},
\label{permit1b}
\end{equation}
such that $\nu_e = \omega_e/(2 \pi \sqrt{\epsilon_0})$ and $\nu_m = \omega_m/(2 \pi \sqrt{\mu_0})$ are the frequencies associated with the electric and magnetic plasmon modes, respectively. We have followed previous work \cite{epsilon-mu} and use $\epsilon_0 = 1.21$ and $\mu_0 = 1.0$ in Eq. (1). Disorder is introduced by allowing the widths of the A and M components at the $j$-th layer to fluctuate around their respective mean values, $a$ and $b$: $a_j=a+\delta^A_j$ and $b_j=b+\delta^M_j$, where the random variables $\delta^{A,M}_j$ are homogeneously distributed in the interval
$[-\Delta/2,\Delta/2]$.
The localization length $\xi$ is calculated numerically using the standard definition~\cite{sheng},
\begin{equation}
\xi^{-1}=-\lim \limits_{L\rightarrow\infty} \left\langle \frac{\ln|T|}{2L}  \right\rangle,
\label{loc definition1}
\end{equation}
where $T$ is the transmission coefficient, and $L$ is the total stack length, $L=\sum_{j=1}^N (a_j+b_j)$, with
$N$ being the total number of double layers;
$\langle \cdots \rangle$ denotes configurational average.

The transmission coefficient is given by \cite{born}
\begin{equation}
 T= {2Z\over Z(M_{22}+M_{11})-Z^2M_{12}-M_{21}},
\label{trans1a}
\end{equation}
where $Z=\cos\theta$, with $\theta$ being the angle of incidence, and
the elements of the transfer matrix $M_{ij}$ are defined as
\begin{equation}
{\bf M}=\begin{bmatrix} M_{11}& M_{12} \\ M_{21}& M_{22}
\end{bmatrix}, \, {\bf M}=\prod\limits_{j=1}^N {\bf M}^{(A)}_j{\bf
M}^{(M)}_j,
\end{equation}
with
\begin{eqnarray}
{\bf M}^{(x)}_j=\begin{bmatrix} \cos q_xx& if_x^{-1}\sin q_xx \\
if_x\sin q_xx & \cos q_xx
\end{bmatrix},\ x=A,M,
\label{mab1}
\end{eqnarray}
$q_x=(\omega/c)u_x(\omega,\theta)$, $u_x(\omega,\theta)\equiv\sqrt{\epsilon_x(\omega)\mu_x(\omega)-\sin^2 \theta}$;
for incident transverse electrical (TE) and transverse magnetic
(TM) waves, the coefficients $f_x$ are
\begin{eqnarray}
f_x^{TE}=\frac{u_x(\omega,\theta)}{\mu_x(\omega)},\
f_x^{TM}=\frac{u_x(\omega,\theta)}{\epsilon_x(\omega)}.
\label{tetm1}
\end{eqnarray}
For an infinitely periodic structure without disorder,
Eqs.\ (\ref{mab1}) and (\ref{tetm1}) lead to the dispersion relation
\begin{equation}
\cos(kd)=\cos(q_Aa)\cos(q_Mb)- {F_+\over 2}
\sin(q_Aa)\sin(q_Mb),
\label{dispersion}
\end{equation}
where $F_\pm=(f_A/ f_M)\pm(f_M/ f_A)$, with $f_x$ corresponding to TM or TE waves; $d=a+b$ is the period of
the system and $k$ is the Bloch wavevector along the
direction of the axis of the periodic photonic crystal.

\begin{figure}[t]
\begin{center}
\includegraphics[ width = 8.9cm,angle=0]{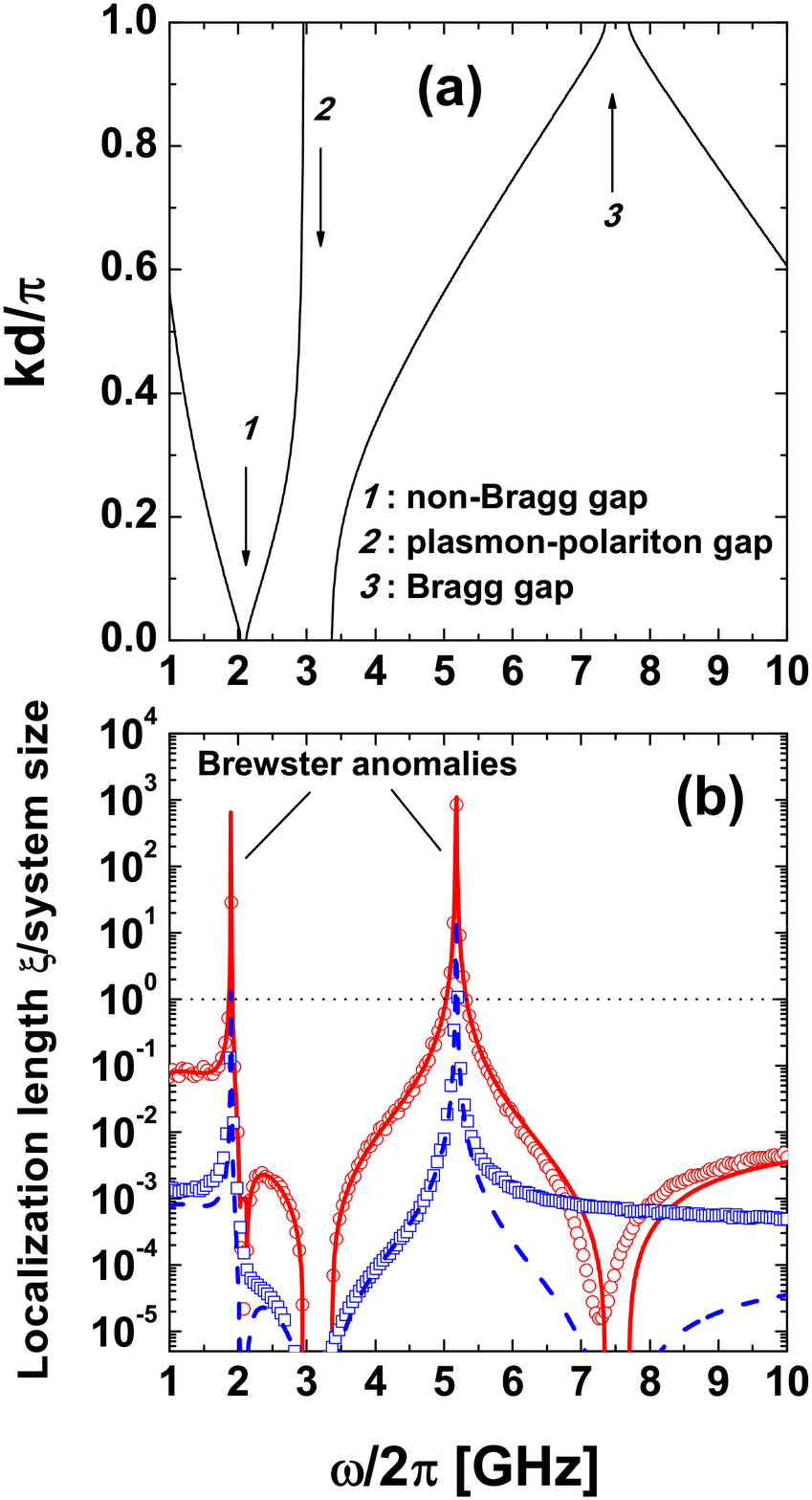}
\end{center}
\caption{(Color online)
(a) Band diagram for TE waves, at incidence angle $\theta=\pi/6$, for a perfectly ordered superlattice, as given by Eq. (\ref{dispersion}).
(b) Localization length $\xi$ (in units of the system size) for TE waves calculated numerically [Eq. (\ref{loc definition1})] for $N=5,000$ double layers and $100$ realizations.
The parameters used are: $a=b=12$ mm, $\omega_m=\omega_e=6\pi$ GHz, $\epsilon_0=1.21$, $\mu_0=1$, and $\Delta=0.5$ mm (open red circles) or $\Delta=12$ mm (open blue squares). The red solid and blue dashed lines correspond to
Eq.~(\ref{approx loc}) for $\Delta=0.5$ mm and 12 mm, respectively.
}
\label{fig1}
\end{figure}

The numerical simulations are supplemented by a generalization (to the case of oblique incidence) of an analytic expression for $\xi$ derived by Izrailev and Makarov (IM) for bilayered photonic structures~\cite{izrailev}, valid for weak disorder (small fluctuating widths in our case),
\begin{equation}
\xi^{-1}={F_-^2\over 8d\sin^2(kd)}[q_A^2\sigma_A^2\sin^2(q_Mb)+
q_M^2\sigma_M^2\sin^2(q_Aa)],
\label{approx loc}
\end{equation}
where $\sigma_x^2=\langle\delta_x^2\rangle$.
For homogeneous random perturbations with the same amplitude on both layers, one has $\sigma_A^2=\sigma_M^2=\Delta^2/12$.

Let us first examine the band diagram for a perfectly periodic structure, as given by Eq. (\ref{dispersion}), and shown in Fig.\ \ref{fig1}(a) for TE waves.
Dispersion leads to the appearance of two non-Bragg gaps: one at a frequency $\omega \simeq 4 \pi$ GHz, corresponding to the vanishing of the average refraction index of the structure, and the other, at $\omega \simeq 6 \pi$ GHz, which only occurs for oblique incidence, $\theta \neq 0$, and is a consequence of the excitation of electric and magnetic (for incident TM and TE waves, respectively) plasmon polaritons~\cite{cavalcanti}.
Figure \ref{fig1}(a) also shows the usual Bragg gap at $\omega \simeq 15 \pi$ GHz.
Figure \ref{fig1}(b) shows the localization length $\xi$ for TE waves as a function of frequency, for $\theta = \pi / 6$.
Before discussing the actual results, two comments are in order: first, we have confirmed that the system is self-averaging, i.e., the behavior of $\xi$ calculated from a single realization of disorder, for a system made up of a sufficiently large number of layers,  does not differ significantly from that obtained through Eq.~(\ref{loc definition1}), considering many disorder realizations for a system not so large;
second, the overall agreement between the simulation results and those from Eq.~(\ref{approx loc}) is excellent for weak disorder (see data for $\Delta=0.5$ mm in Fig.~\ref{fig1}), but deviations occur as disorder increases, as expected, as illustrated in Fig.~\ref{fig1}(b) for $\Delta=12$ mm. A comparison between Figs.\ \ref{fig1}(a) and (b) indicates that for weak disorder the dips in $\xi$ correlate with Bragg and non-Bragg gaps alike; however, as the strength of disorder increases, the Bragg gap at $\omega \simeq 15 \pi$ GHz is washed out, leading to the disappearance of the associated dip in $\xi$.
The TM waves follow the same pattern.

\begin{figure}[t] 
\begin{center}
\includegraphics[ width = 5.8cm,angle=0]{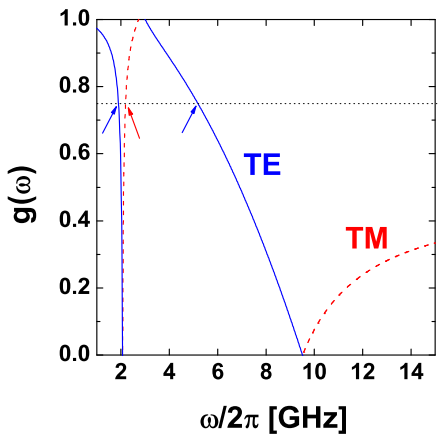}
\end{center}
\caption{(color online) The function $g(\omega)$, given by the RHS of Eq.~(\ref{brewster}), as a function of frequency for TE (full curve) and TM (dashed curve) waves. For a given angle of incidence, $\theta$, the Brewster anomalies occur at frequencies satisfying $\cos^2\theta=g(\omega)$. As an example, the horizontal dotted line corresponds to $\cos^2\pi/6$, whose intersections (indicated by arrows) with $g(\omega)$ yield the frequencies for Brewster anomalies for $\theta = \pi/6$. The media parameters are the same as in Fig.~\ref{fig1}.
\color{red}
\color{black}
}
\label{fig2}
\end{figure}

Most importantly, Fig.~\ref{fig1}(b) shows that there are some specific frequencies at which $\xi$ reaches anomalously high values, larger than the system size, hence leading to light delocalization.
This is a manifestation of the ``Brewster anomaly"~\cite{sipe1988}, corresponding to the situation in which no reflection occurs in 1D disordered optical systems at specific incident angles (Brewster angles, $\theta_B$).
To see how this comes about, one first considers non-magnetic materials ($\mu_M = 1$), and a TM wave incident on an interface at an angle $\theta_B$: there is no reflected component, since the induced electric field cannot radiate along its own axis. Multiple reflections are therefore eliminated along the medium, and localization is suppressed.
For the metamaterials considered here ($\mu_M,\epsilon_M \neq 1$), the reflected field is a result of radiation from both electric and magnetic dipoles, thus explaining the presence of sharp peaks in $\xi$ both for TM and TE waves; Brewster anomalies for TM and TE waves have been observed at RH-LH interfaces~\cite{Tamayama06}.
It is important to remark that the Brewster anomaly may occur at the vicinity of a band gap, so that the presence of disorder does not halt light propagation at band edges.
Furthermore, at the band edge, group velocities may be very small and one could observe slow light propagation.

\begin{figure} [t] 
\begin{center}
\includegraphics[ width = 7.6cm,angle=0]{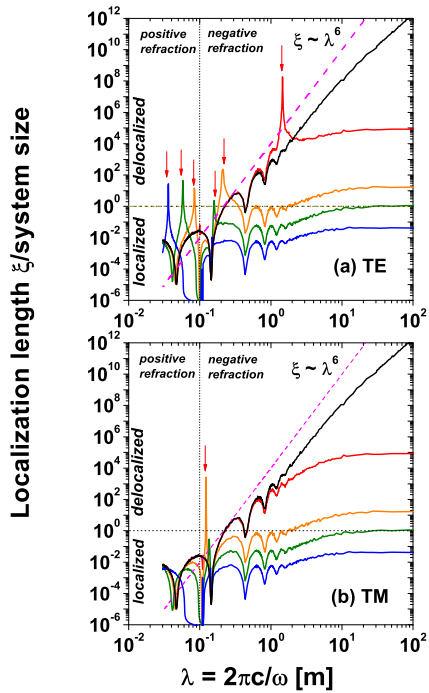}
\end{center}
\caption{(color online) Localization length $\xi$ (in units of the system size) for TE (a) and TM (b) waves as a function of the vacuum wavelength $\lambda$, obtained from our numerical simulations, for different angles of incidence (solid lines, from top to bottom): $\theta = 0$, $\pi/100$, $\pi/12$, $\pi/6$, and $\pi/3$. Vertical arrows locate the Brewster modes, and the dashed line corresponds to the asymptotic behavior $\xi \propto \lambda^{6}$ predicted in Ref.~\cite{asatryan}. The media parameters are the same as in Fig.~\ref{fig1}, with $\Delta=0.5$ mm.
\color{red}
\color{black}
}
\label{fig3}
\end{figure}

Further insight into the frequency dependence of the Brewster angles can be obtained by examining the conditions for their occurrence within the framework of the generalized IM theory.
According to Eq.\ (\ref{approx loc}), $\xi$ diverges when $F_-\to 0$, which occurs for values of $\theta$ such that
\begin{equation}
\cos^2 \theta=g(\omega)\equiv{\epsilon_M(\omega)\mu_M(\omega)-1\over p^2(\omega)-1},
\label{brewster}
\end{equation}
where $p(\omega)=\mu_M(\omega)$ for TE waves, and $p(\omega)=\epsilon_M(\omega)$ for TM waves.
To illustrate this condition, Fig.~\ref{fig2} shows the frequency dependence of the Brewster angles, as calculated from Eq.~(\ref{brewster}), with the dispersive relations given by Eq.\ (\ref{permit1b}), for both TM and TE configurations.
One sees that, as $\theta$ increases, one goes from a regime in which the anomalies are present for three frequencies (corresponding to two TE and one TM mode, c.f. Fig.~\ref{fig2} for $\theta=\pi/6$) to one in which an additional TM mode appears.
Equation~(\ref{brewster}) also puts in evidence that it is crucial to take dispersion into account in order to determine the values of Brewster angles in metamaterials.

Let us now discuss the behavior of the localization length over a wide range of frequencies, for different angles of incidence, $\theta$.
Figure~\ref{fig3} shows the simulation results for $\xi$, for the specified set of parameters; from now on we refer to $2\pi c/\omega$ as the (vacuum) wavelength $\lambda$. First we note that, for normal incidence, $\xi\propto \lambda^6$, similar to the behavior found in Ref.\ \cite{asatryan}.
However, this power-law behavior is already lost for very small angles of incidence (e.g., for $\theta=\pi/100$); nevertheless, $\xi$ may be larger than the system size in the long wavelength limit for small angles, $\theta\lesssim \pi/12$, thus suppressing localization. The long-wavelength delocalization for normal incidence is due to the fact that we assumed $\omega_e=\omega_m$.
Indeed, the low-frequency limit of the diverging condition, Eq.~(\ref{brewster}), yields $\cos \theta \sim \omega_e/\omega_m$ for TE waves ($\omega_m/\omega_e$ for TM waves), so that $\theta=0$; by the same token, for $\omega_e\neq\omega_m$, light delocalization in the low-frequency regime occurs at $\theta\simeq\cos^{-1} [\omega_e/\omega_m]^{\pm 1}$, which highlights the importance of allowing for oblique incidence. For larger angles ($\theta \gtrsim \pi/12$), the only mechanism for suppression of localization corresponds to the Brewster anomalies at well defined wavelengths: for TE modes and for a given angle of incidence, localization is suppressed at two different wavelengths, one in the negative- and one in the positive-refraction regions of the metamaterial; for the TM modes, as discussed in connection with Fig.~\ref{fig2}, one must reach a critical angle of incidence before a Brewster anomaly emerges in the positive-refraction regime.

Finally, as dispersive metamaterials are intrinsically absorptive, comments regarding losses are in order. The resonant nature of the electromagnetic response of the metamaterial, notably the magnetic one, is the main source of dissipation~\cite{shalaev09}; other sources of dissipation, such as surface roughness and interface effects, do exist and constitute a major hurdle in the development of metamaterials. In the present Drude-like model, absorption may be introduced phenomenologically through the replacement $\omega^2\to\omega(\omega+i\gamma_{e,m})$, where $\gamma_{e,m}$ are the electric and magnetic loss factors; we have found that the Brewster modes are indeed smeared out when absorption effects are included. However, according to Eq.~(\ref{brewster}), one may conceive realizing a metamaterial with dielectric and magnetic responses such that Brewster modes would show up in a spectral region in which losses are not too severe; in so doing, they could still lead to observation of light delocalization. Moreover, important progress in mitigating losses in metamaterials has been achieved recently, such as incorporation of gain~\cite{gain}, and the use of optical-parametric amplification~\cite{shalaev09,opa}, which has the advantage of being tunable in a wide negative index frequency range. These advances have led to significant attenuation and complete compensation of losses; even lasing in metamaterials has been observed~\cite{lasing}. Therefore, a combination of these loss-control mechanisms with the routes proposed here should lead to a wide range of possibilities for control of light flow and Anderson localization in disordered metamaterials.

In conclusion, we have studied light propagation through 1D disordered structures composed of alternating layers of non-dispersive and dispersive metamaterials; the disorder consists in allowing for randomness in the layer widths.
We have found that the inclusion of dispersion, together with the possibility of oblique incidence, lead to novel routes for suppression of Anderson localization of light, in addition to the long wavelength limit for normal incidence~\cite{asatryan}. First, for a given choice of dispersive metamaterials, oblique incidence at $\theta\simeq\cos^{-1} [\omega_e/\omega_m]^{\pm 1}$ leads to delocalization in the long wavelength limit. The second route is provided by the Brewster anomalies, corresponding to a diverging localization length at well defined frequencies. Interestingly, these anomalies can occur for frequencies in ranges such that the metamaterial exhibits either negative- or positive refraction; thus, negative refraction is not a necessary condition for delocalization, but the use of dispersive materials is.
These results indicate that control and fine-tuning of multiple scattering of light could be achieved by a careful combination of angle of incidence and a  judicious choice of metamaterials.

\begin{acknowledgments}
Financial support from the Brazilian Agencies CNPq, CAPES, FAPERJ, FAPESP, and FUJB is gratefully acknowledged.
\end{acknowledgments}

\end{document}